\documentclass{LMCS}
\pdfoutput=1
\usepackage{amsmath}
\usepackage{amssymb}
\usepackage{meq}
\usepackage{xspace}
\usepackage{bussproofs}
\EnableBpAbbreviations
\usepackage{haskell}
\usepackage{hyperref}
\usepackage{header}
\usepackage{framed}
\usepackage{wrapfloat}
\usepackage{diagrams}
\usepackage{enumerate}

\DeclareMathAlphabet{\mathsfbf}{OT1}{cmss}{bx}{n}

\def\doi{6 (3:15) 2010}
\lmcsheading%
{\doi}
{1--23}
{}
{}
{Nov.~13, 2009}
{Sep.~\phantom03, 2010}
{}

\begin{document}

\title{Initial Algebra Semantics for Cyclic Sharing Tree Structures\rsuper*}

\author[M.~Hamana]{Makoto Hamana}
\address{Department of Computer Science, Gunma University, Japan}
  \email{hamana@cs.gunma-u.ac.jp}

\titlecomment{{\lsuper*}Some of the results presented here were first published in TLCA
proceedings \cite{tlca}.
}

\begin{abstract}

Terms are a concise representation of tree structures. Since they
can be naturally defined by an inductive type, they offer data
structures in functional programming and mechanised reasoning with
useful principles such as structural induction and structural
recursion. However, for graphs or "tree-like" structures --
trees involving cycles and sharing -- it remains unclear
what kind of inductive structures exists and how we can faithfully
assign a term representation of them.  In this paper we propose a
simple term syntax for cyclic sharing structures that admits
structural induction and recursion principles. We show that the
obtained syntax is directly usable in the functional language
Haskell and the proof assistant Agda, as well as ordinary data
structures such as lists and trees.  To achieve this goal, we use
a categorical approach to initial algebra semantics in a presheaf
category. That approach follows the line of Fiore, Plotkin and
Turi's models of abstract syntax with variable binding.

\end{abstract}



\maketitle

\section{Introduction}
\W{Terms} are a convenient, concise and mathematically clean
representation of tree structures
used in logic and theoretical computer science.
In the fields of traditional algorithms and
graph theory, one usually uses 
unstructured representations for trees,
such as pairs $(V,E)$ of vertices and edges sets, 
adjacency lists, adjacency matrices, pointer structures, etc.
Such representations are
more complex and unreadable than terms.
We know that term representation provides a well-structured, compact and 
more readable notation.

However, consider the case of a ``tree-like'' structure such as 
that depicted below.  
\begin{wrapfigure}[7]{l}{12em}
\y{-.9em}
\includegraphics[scale=.28]{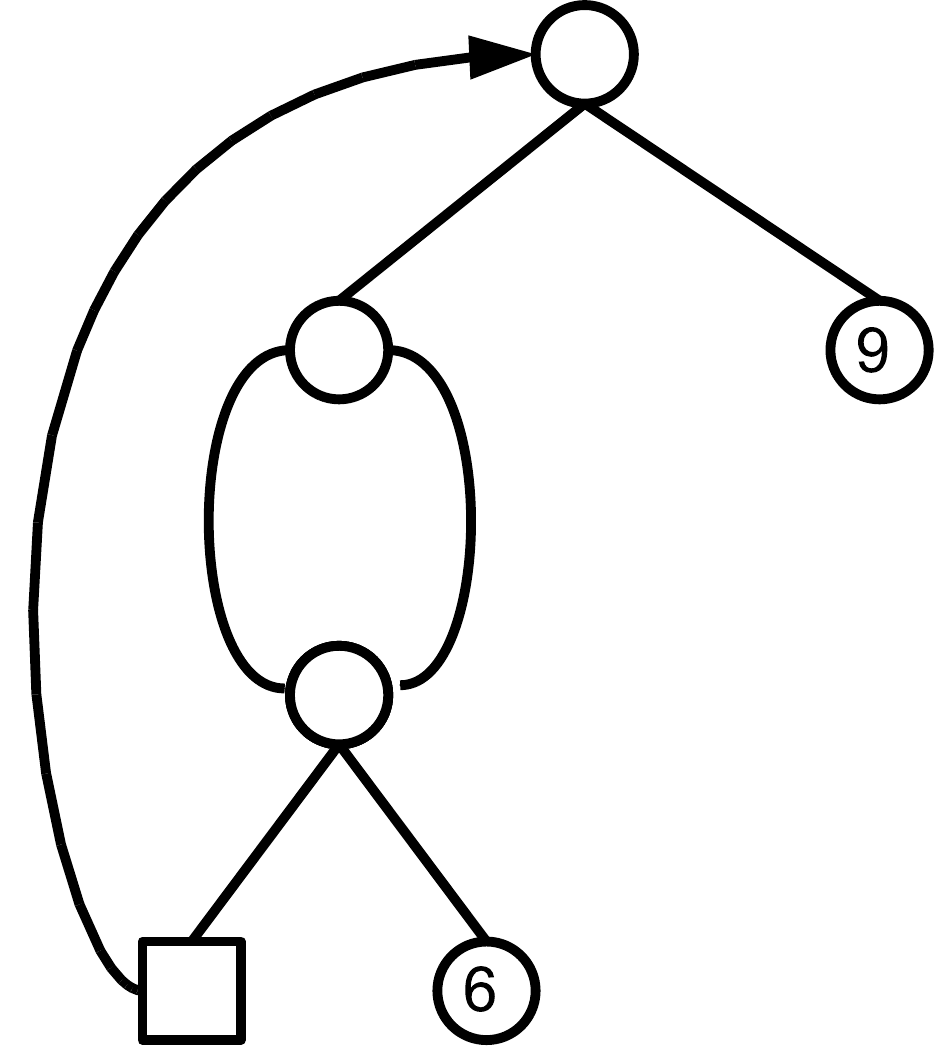}
\label{fig:cstree2n}
\end{wrapfigure}
This kind of structure -- graphs, but almost trees involving (a few) 
exceptional edges --
quite often appears in logic and computer science.
Examples include internal representations of expressions 
in implementations of functional languages 
that share common sub-expressions for 
efficiency, 
data models of XML such 
as trees with pointers \cite{contextlogic}, proof trees admitting cycles
for cyclic proofs \cite{Brotherston:05}, 
term graphs in graph rewriting \cite{Baren,ETGRS}, and 
control flow graphs of imperative programs 
used in static analysis and compiler optimizations \cite{dominator}.
Suppose that we want to treat such structures in a pure functional programming
language such as Haskell, Clean,
or a proof assistant such as Coq, Agda \cite{Agda}.
In such a case, we would have to
abandon the use of naive term representation, 
and
would instead be compelled to
use an unstructured representation such as $(V, E)$,
adjacency lists, etc.
Furthermore, a serious problem
is that we would have to
abandon structural recursion and induction to decompose them
because they look ``tree-like'' but are in fact graphs, so
there is no obvious inductive structure in them.
This means that in functional programming,
we cannot use pattern matching to treat tree-like structures,
which greatly decreases their convenience.
This lack of structural induction implies a
failure of being an inductive type.
%
%
But,
are there really no inductive structures in
tree-like structures?
As might be readily apparent, 
tree-like structures are almost trees and merely contain finite pieces of
information. The only differences are the presence of ``cycles''
and ``sharing''.

In this paper, we give an initial algebra characterisation of
cyclic sharing tree structures
in the framework of categorical universal algebra.
%
The aim of this paper is to derive the following practical goals 
\W{from the initial algebra characterisation}.

\begin{enumerate}[{[I]}]
\item To develop a simple term syntax
for cyclic sharing structures
that admits structural induction and structural recursion principles.
\label{item:ind}
\item To make the obtained syntax directly
usable in the current functional languages and proof assistants,
as well as ordinary data structures, such as lists and
trees.
\label{item:impl}
\end{enumerate}
The goal [\ref{item:ind}] requires that the term syntax \W{exactly}
represents cyclic sharing structures (i.e. no junk terms exist)
to make structural induction possible.
The goal [\ref{item:impl}]
requires that the obtained syntax
should be \W{lightweight as possible},
which means that e.g. well-formedness and 
equality tests on terms for cyclic sharing
structures
should be fast and easy, as are ordinary data structures
such as lists and trees.
We do not want many axioms to characterise the intended structures,
because, in programming situation,
checking the validity of axioms every time
is expensive and makes everything complicated.
Ideally, formulating structures \W{without axioms} is best.
Therefore, 
the goal [\ref{item:impl}] is rephrased more specifically as:
\begin{enumerate}[{[I']}]
\setcounter{enumi}{1}
\item To give an inductive type
that represents cyclic sharing structures \W{uniquely}. 
We therefore rely on that
a type checker automatically ensures the well-formedness of cyclic sharing
structures. \label{item:inddata}
\end{enumerate}

\noindent To show this,
we give concrete definitions of
types for cyclic sharing structures
in two systems: a functional programming language
Haskell and a proof assistant Agda.

\subsection{Variations on initial algebra semantics}
The initial algebra semantics models 
syntax/datatype as the initial algebra and semantics
as another algebra, and the compositional interpretation
by the unique homomorphism.
The classical  formulation of
initial algebra semantics for syntax/datatype taken by ADJ \cite{ADJ}
is categorically reformulated as
an initial algebra of a functor in the category
\Set of sets and functions
\cite{Robinson}, which means that carriers are merely sets and 
operations are functions.

Recently, varying the base category other than \Set,
initial algebra semantics for algebras of functors
has proved to be
useful framework for characterisation of various mathematical
and computational structures
in a uniform setting. We list several:
%
%
$S$-sorted abstract syntax is characterised as initial algebra in $\Set^S$
 \cite{Robinson},
second-order abstract syntax as initial algebra 
in \SetF \cite{FPT,free,
Fiore2nd} (where 
\FF is the category of finite sets),
explicit substitutions as initial algebras
in the category $[\Set,\Set]_f$ of finitary functors \cite{ExHOSC},
recursive path ordering for term rewriting systems
as algebras in the category \cat{LO} of linear orders \cite{Ryu},
second-order rewriting systems as initial algebras
in the preorder-valued functor category $\cat{Pre}^\FF$ \cite{CRS}, 
and
nested datatypes \cite{ghani-tlca07}  and
generalised algebraic datatypes (GADTs) \cite{NeilGADT}
in functional programming
as initial algebras
in $[\CC, \CC]$ and $[|\CC|, \CC]$, respectively, where \CC is a 
\omega-cocomplete category.

This paper adds a further example to the list given above.
We characterise cyclic sharing structures as 
an initial algebra in the category \SetFTT, where \TT is 
the set of all ``shapes'' of trees and
\FT is the set of all tree shape contexts.
We derive
structural induction and recursion principles from it.
An important point 
is that
we merely use \W{algebra of functor}
to formulate cyclic sharing structures, i.e. 
not (models of) equational specifications or 
$(\Sig,E)$-algebras.
This characterisation achieves 
the requirement of ``without axioms''. Moreover, it is the key to
formulate them by an inductive type.



\subsection{Basic idea}
It is known in the field of graph algorithms \cite{Tarjan} that, 
by traversing a rooted directed graph in a depth-first search manner,
we obtain a \W{depth-first search tree}, which
consists of a spanning tree (whose edges are called \W{tree edges}) 
of the graph and 
\W{forward edges} (which connect ancestors in the tree to descendants), 
\W{back edges} (the reverse), and \W{cross edges} 
(which connect nodes across the tree from right to left).

\begin{wrapfigure}[10]{l}{12em}
\includegraphics[scale=.27]{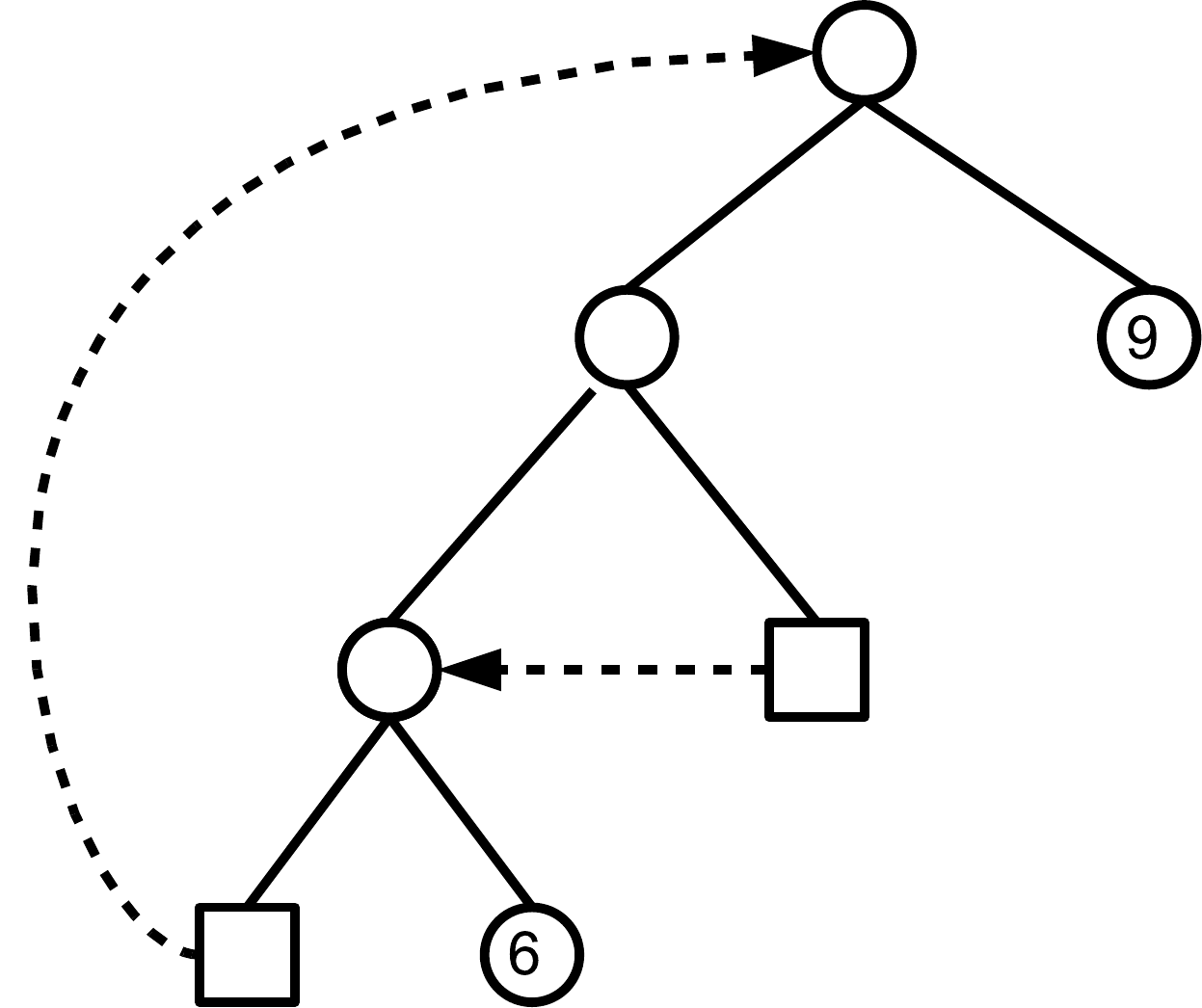}
\caption{Depth-first search tree}\label{fig:cstree4}
\end{wrapfigure}
\noindent
Forward edges can be decomposed into
tree and cross edges by placing indirect nodes.
For example, the graph in the front page 
becomes a depth-first search tree in Fig.\ref{fig:cstree4}
where solid lines are tree edges and dashed lines are 
back and cross edges.
This is the target structure we will model in this paper. 
That is, tree edges are the basis of an \W{inductive structure},
back edges are used to form \W{cycles},
and cross edges are used to form \W{sharing}.
Consequently, our task is to seek how to characterise 
pointers that make back edges and cross edges in inductive constructions.

\subsection{Formulation}
The crucial idea to formulate pointers in inductive constructions
is to use \W{binders} as \W{pointers} in
abstract syntax.
Trees are formulated as terms. 
Hence, a remaining problem is
how to exactly capture binders in terms.
Fiore, Plotkin and Turi \cite{FPT} have characterised
abstract syntax with variable binding
by initial algebras in the presheaf category \SetF,
where \FF is the category of finite sets.
For example, abstract syntax of \lmd-terms is modeled as a functor
\[
\Lambda : \FF \rTo \Set
\]
equipped with three constructors for \lmd-terms as an algebra structure on \Lambda.
Each set 
$\Lambda(X)$ gives the set of all \lmd-terms which may contain free variables
taken from a set $X$ in \FF.
This formulation models a structure (here,
abstract syntax trees)
indexed by suitable invariant (here, free variables
considered as contexts), 
which is essential information
to capture the intended structure (abstract syntax with variable
binding).

However, this approach using algebras in \SetF
is insufficient to represent ``cross edges'' in 
tree-like graphs.
Ariola and Klop \cite{ETGRS} have analysed that
there are two kinds of sharing in this kind of
tree-like graphs:
\begin{enumerate}[(i)]
\item {vertical} sharing (i.e. back edges in depth-first search trees), and
\item {horizontal} sharing (i.e. cross edges).
\end{enumerate}
In principle, binders capture ``vertical'' contexts only,
but to represent cross edges exactly,
we must capture ``horizontal'' context information
that cannot be handled by the index category \FF.
%

To solve this problem, in this paper we take
a richer index category that is enough to model 
cross edges.
We introduce the notion of \W{shape trees} 
and contexts consisting of them, which represents
other parts of tree viewing from a pointer node.
We use the set \TT of all shape trees as ``types'' of syntax,
and the set \FT of all sequences of shape trees as ``context''.
We follow
Fiore's treatment of initial algebra 
semantics for typed abstract syntax with variable binding \cite{NBE}
by
algebras in the presheaf category
$\SetFTT$. 
Therefore, cyclic sharing trees are modelled as a 
\TT and \FT-indexed set
\[
T :  \TT  \rTo (\FT \rTo \Set)
\]
equipped with constructors of cyclic sharing trees as an algebra structure.

\subsection{Organisation} 
We first give types and abstract syntax
for cyclic sharing binary trees in Section \ref{sec:share}.
We then characterise cyclic sharing binary trees as an initial algebra
in Section \ref{sec:ini}.
Section \ref{sec:impl} gives a way of implementing cyclic sharing structures 
by inductive types.
Section \ref{sec:gen} generalises our treatment to arbitrary signature
for cyclic sharing structures.
Section \ref{sec:vari} presents discussion of
variations of the form of pointers in cyclic sharing trees.
Section \ref{sec:etg} relates our representation
and equational term graphs
in the initial algebra framework by giving a homomorphic translation.
In Section \ref{sec:othersem}, we discuss connections to other approaches
to cyclic sharing structures.


\section{Abstract Syntax for Cyclic Sharing Structures}


\subsection{Cyclic structures by \mu-terms}\label{sec:cyc}
The \mu-notation (\mu-terms) for fixed point expressions is widely
used in computer science and logic.
Its theory has been investigated thoroughly, 
for example, in \cite{ETGRS,alex-plot}.
The language of \mu-terms suffices to express all
cyclic structures. 

\begin{figure}[t]
\begin{center}
\includegraphics[scale=.35]{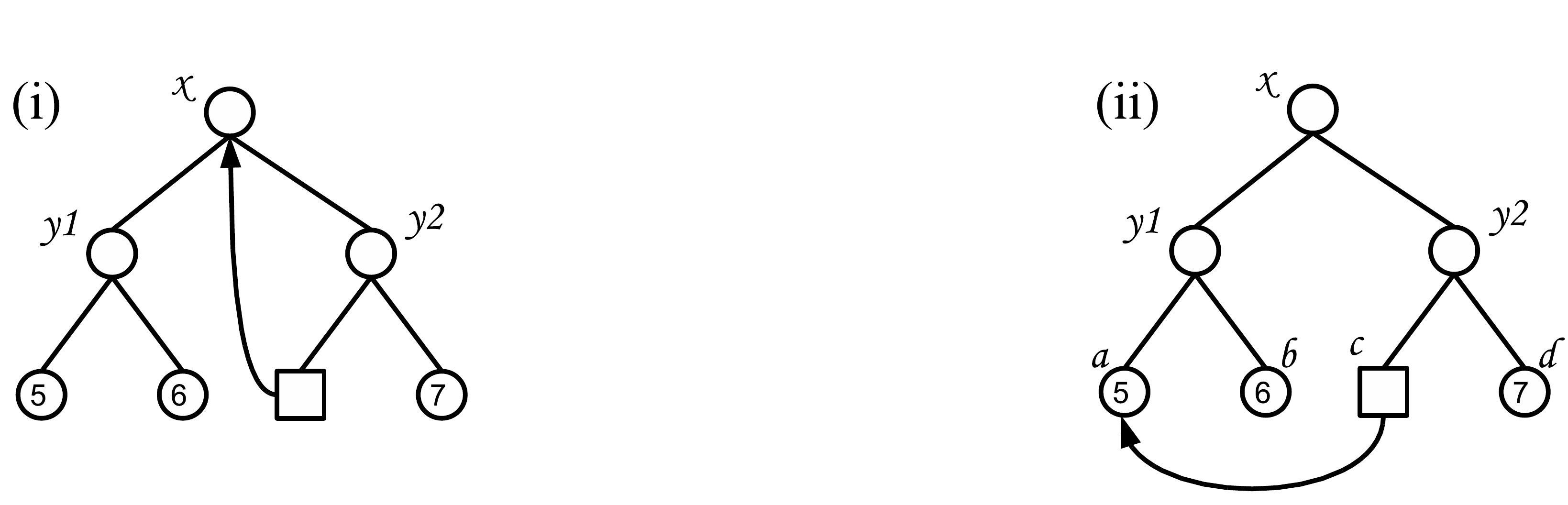}
\y{-1em}
\caption{Trees involving cycle and sharing}\label{fig:cstree7}
\end{center}
\hrule height 0.5pt
\end{figure}

For example, a cyclic binary tree shown in
Fig. \ref{fig:cstree7} (i) is representable as
the term
\begin{equation}
\mu x . 
  \bin( \mu y_1 . \bin(\lf( 5),\lf( 6)),\,
        \mu y_2 . \bin(\,x,\, \lf( 7) \,))
\label{eq:cyc}
\end{equation}
where \bin and \lf respectively denote a binary node and a leaf.
The point is that
the variable $x$ refers to the root labeled by a \mu-binder,
hence a cycle is represented.
To uniquely formulate cyclic structures, here 
we introduce the following assumption:
we attach \mu-binders in front of \bin only,
and put exactly one \mu-binder for each occurrence of \bin as for (\ref{eq:cyc}).
This is seen as uniform addressing of \bin-node, i.e.,
$x,y_1,y_2$ are seen as labels or ``addresses'' of \bin-nodes.
We also assume no axiom to equate \mu-terms.
That is, we do not identify a \mu-term with
its unfolding,
since they are different (shapes of) graphs.
In summary, \mu-terms
represent cyclic structures.
This is the underlying idea of a representation of
cyclic data given in \cite{TFPcyc}
by using the functional programming language Haskell.


\subsection{How to represent sharing}
\label{sec:share}
Next, we incorporate sharing.
The presence of sharing makes the
situation more difficult.
Consider the tree (ii) in Fig. \ref{fig:cstree7}
involving sharing via
a cross edge.
As similar to the case of cycles,
this might be written as a \mu-term
\[
\mu x . 
  \bin( \mu y_1 . \bin(\lf( 5),\lf( 6)),\,
        \mu y_2 . \bin(\,\RM{\fbox{\phantom{$\ptr{11}{x}$}}}\,,\lf( 7))).
\]
But can we fill the blank to refer the node $a$
(in Fig. \ref{fig:cstree7} (ii))
from the node $c$ ``horizontally'' by
using the mechanism of binders?
Actually, \mu-binders are insufficient for this purpose.
Therefore, we introduce
a new notation ``$\ptr{p}{x}$''
to refer to a node horizontally.
This notation means going up to a node $x$ labelled by a \mu-binder
and going down to a position $p$ in the
subtree rooted by the node $x$.
In the example presented above, 
the blank is filled as
\[
\mu x . 
  \bin( \mu y_1 . \bin(\lf( 5),\lf( 6)),\,
        \mu y_2 . \bin(\,\RM{{{$\ptr{11}{x}$}}}\,,\lf( 7))).
\]
The pointer $\ptr{11}{x}$
means going back to the node $x$, then
going down through the left child twice (using the position $11$).
See also Example \ref{ex:left}. 
In this section, we focus on the formulation of
binary trees involving cycles and sharing.
Binary trees are the minimal case that can involve 
the notion of sharing in structures. 
Later, in Section \ref{sec:gen}, we will consider general data types.


\subsection{Shape trees}
We designate our target data structures as \W{cyclic sharing trees} and
its syntactic representation \W{cyclic sharing terms}.
Cyclic sharing trees are binary trees 
generated by nodes of three kinds, i.e., pointer, leaf, and binary node,
and satisfying a certain condition of well-formedness.

To ensure correct sharing, 
we introduce the notion of \W{shape trees},
which are skeletons of cyclic sharing trees.
That is, shape trees are binary trees, forgetting values in pointer nodes and 
leaves from cyclic sharing trees. 
The set \TT of all shape trees is defined by
\[
\TT \ni \tau ::=  \svoid  \| \sptr \| \slf \| \sbin(\tau_1,\tau_2)
\]
where
\svoid is the void shape,
\sptr is the pointer node shape,
\slf is the leaf node shape, and
 $\sbin(\tau_1,\tau_2)$ is 
the binary node shape.
We typically use Greek letters $\sig, \tau$ to denote shape trees.

We define referable positions in a shape tree.
A \W{position} is a finite sequence of $\set{1,2}$.
The root position is denoted by the empty sequence \epsilon and 
the concatenation of positions is denoted by
$p  q$ or $p.q$.
The set $\Pos(\tau)$ of referable
positions in a shape tree $\tau$ is defined by
\begin{meqa}
\Pos(\svoid) &= \Pos(\sptr) = \emptyset\\
\Pos(\slf) &= \set{\epsilon}\\
\Pos(\sbin(\sig,\tau)) &= \set{\epsilon} \union 
\set{1  p \| p \in \Pos(\sig)}\union
\set{2  p \| p \in \Pos(\tau)}.
\end{meqa}
An important point is
that the void \svoid and the pointer \sptr nodes
are not referable by other nodes,
hence their positions are defined to be empty sets.

\subsection{Syntax and types}
\label{sec:typerule-share}
Shape trees are used as types in a typing judgment.
As usual, a typing context \Gamma is a sequence of 
(variable, shape tree)-pairs.
\begin{framed}
\x{-1em}\textbf{Typing rules}
\begin{meq}
\infrule{Pointer}
{p\in \Pos(\sig)}
 {}
{\Gamma,x:\sig,\Gamma' \pr \ptr{p}{x} : \sptr}
\quad
\infrule{Leaf}
{k \in \ZZ}
 {}
{\Gamma \pr \lf( k) :  \slf} 
\\[.5em]
\infrule{Node}
{x:\sbin(\svoid,\svoid), \Gamma  \pr s : \sig \infspc 
 x:\sbin(\sig,\svoid), \Gamma  \pr t : \tau}
 {}
{\Gamma \pr \mu x.\bin(s,t) :   \sbin(\sig,\tau)}
\end{meq}
\end{framed}

In these typing rules,
a shape tree type is assigned to the corresponding tree node.
That is, a binary node is of type $\sbin(\sig,\tau)$
of binary node shape, a pointer node is of type $\sptr$
of pointer node shape, and a leaf node is of type $\slf$ of leaf node shape.

A type declaration $x \!:\!\sig$ in a typing context (roughly) means that
\sig is the shape of subtree (say, $t$) headed by a binder $\mu x$
(see Example \ref{ex:left}). Hence, in (Pointer) rule,
taking a position $p\in\Pos(\sig)$, we safely refer to 
a position in the tree $t$.
The notation $\ptr{p}{x}$
is designed to realise a right-to-left cross edge.
Note also that a path obtained by $\ptr{p}{x}$
is the shortest path from the pointer node to the node referred 
by $\ptr{p}{x}$.
When $p=\epsilon$,
we abbreviate $\ptr{\epsilon}{x}$ as $\upptr{x}$.
This $\upptr{x}$ exactly expresses a back edge.
In (Node) rule, the shape trees $\sbin(\svoid,\svoid)$ and 
$\sbin(\sig,\svoid)$ mask nodes that are reachable via 
left-to-right references (i.e. not our requirement) 
or redundant references (e.g. going up to a node $x$ then going back down 
through the same path)
by the void shape \svoid.


\begin{eExample}\label{ex:left}
The binary tree involving sharing in Fig. \ref{fig:cstree7} (ii)
is represented as a well-typed term
\[
\mu x . 
  \bin( \mu y_1 . \bin(\lf( 5),\lf( 6)),\,
        \mu y_2 . \bin(\ptr{11}{x},\lf( 7))).
\]

\medskip
\noindent
Its typing derivation is the following.

\medskip
\def\defaultHypSeparation{\hskip 0in}
\newcommand{\prftreeA}{
\AXC{$
y_1\: \alpha,
x\:  \alpha
  \pr \lf(5) : \slf
$}
\AXC{$
y_1\:  \sbin(\slf,\svoid),
x\:  \alpha
  \pr \lf(6) : \slf
$}
\BinaryInfC{$
x\:  \alpha
  \pr \mu y_1.\bin(\lf(5),\lf(6)) : \sbin(\slf,\slf)
$}
}

{\small
\begin{prooftree}
\prftreeA
\AXC{$
11 \in \Pos(\beta)
$}
\UnaryInfC{$
y_2\:  \alpha,
x\: \beta
\pr \ptr{11}{x}: \sptr
$}
\AXC{$
y_2\: \sbin(\sptr,\svoid),
x\: \beta
\pr \lf(7) : \slf
$}
\BinaryInfC{$
x\:\beta
  \pr \mu y_2.\bin(\ptr{11}{x},\lf(7)) :  \sbin(\sptr,\slf)
$}
\BinaryInfC{$
\pr \mu x. \bin(
\mu y_1.\bin(\lf(5,\lf(6)),
\mu y_2.\bin(\ptr{11}{x},\lf(7))
) :
 \sbin(
  \sbin(\slf,\slf),
  \sbin(\sptr,\slf))
$}
\end{prooftree}
}
%



\bigskip

\noindent
where $\alpha=\sbin(\svoid,\svoid),\; \beta = \sbin(\sbin(\slf,\slf),\svoid)$.

\end{eExample}

As a result, we can ensure that
no dangling pointer happens in this type system.

\begin{theorem}[Safety]\label{th:safe}
If a closed term $\pr t : \tau$ is derivable,
any pointer in $t$ points to a node in $t$.
\end{theorem}
\proof
By the typing rules,
it is obvious that
a variable $x$ in a pointer in the resulting $t$
is always taken from a \mu-binder in $t$.
Looking at (Node) rule from the lower to the upper,
the shape $\sbin(\sig,\tau)$ is always decomposed, and
two shape trees in typing contexts at the upper
contain fewer possible positions than the lower, i.e.
\[
\Pos(\sbin(\svoid,\svoid)) \subseteq
\Pos(\sbin(\sig,\svoid)) \subseteq
\Pos(\sbin(\sig,\tau)).
\]
This means that  
at any application of (Pointer) rule at the topmost of
a typing derivation tree, 
a taken position $p$ is included in 
the positions of a sub-shape tree of the type of $t$
at the bottom of the typing derivation.
\qed

\subsection{De Bruijn version}
Instead of named variables for binders,
a de Bruijn notation is also possible.
The construction rules are reformulated as follows.
Now a typing context \Gamma is simply a sequence of 
shape trees $\tau_1,\ooo,\tau_n$.
Let $|\Gamma|$ denote its length.
A judgment $\Gamma \pr t : \tau$ denotes a well-formed term $t$
of shape \tau
containing free variables (de Bruijn indices)
from $1$ to $|\Gamma|$.
The intended meaning is that
the length $|\Gamma|$ denotes how many maximally we can go up 
from the current node $t$, and
each shape tree $\tau_i$ in \Gamma denotes the shape of the 
subtree at $i$-th upped node from $t$.
Consequently, when $t$ is a pointer, a context specifies
the set of all positions to which a pointer node can refer.

As known from \lmd-calculus, using de Bruijn notation,
binders become nameless.
Therefore we can safely omit ``$x$'' from $\mu x$.
Because the typing rules are designed to
attach exactly one \mu-binder for each \bin,
even ``\mu'' can be omitted. 
As a result, we obtain a simplified construction rules of terms.
%
\begin{framed}
\x{-1em}\textbf{Typing rules (de Bruijn version)}
\begin{meq}
\infrule{dbPointer}
{|\Gamma|=i-1 \infspc p \in \Pos(\sig)}
 {}
{\Gamma,\sig,\Gamma' \pr \ptr{p}{i} : \sptr}
\quad
\infrule{dbLeaf}
{k \in \ZZ}
 {}
{\Gamma \pr \lf( k) :  \slf} 
\\[.5em]
\infrule{dbNode}
{\sbin(\svoid,\svoid), \Gamma  \pr s : \sig \infspc 
 \sbin(\sig,\svoid), \Gamma  \pr t : \tau}
 {}
{\Gamma \pr \bin(s,t) :   \sbin(\sig,\tau)}
\end{meq}
\end{framed}

In the (dbPointer) rule, 
the condition $|\Gamma|=i-1$
states that the shape tree
\sig appears at $i$-th position of the typing context in
the lower judgment.
Because its depth-first search tree is unique
for a given graph,
the following is immediate.

\begin{theorem}[Uniqueness]
Given a rooted graph that is connected, directed and edge-ordered
with each node having out-degree at most $2$,
the term representation in de Bruijn is unique.
\end{theorem}

\begin{eRemark}
This uniqueness of term representation has
practical importance.  For instance, for the graph in
the tree (ii) in Fig. \ref{fig:cstree7},
there is only one way to represent it in this
term syntax, i.e., $\bin( \bin(\lf(5,\lf(6)), \bin(\ptr{11}{2},\lf(7)) )$
in de Bruijn.  
Therefore, we do not
need any complex equality on graphs (other than the syntactic equality)
to check whether given data are the required data. 
This contrasts directly to other approaches.
If we represent a graph as
a term graph with labels \cite{Baren}, an equational term graph \cite{ETGRS}, or
a \sLETREC-term \cite{Hassei}, 
then several syntactic representations exist
for a single graph. 
Therefore, some normalisation is required, 
for instance when
defining
a function on graphs.
Generally speaking, our terms
are regarded as ``de Bruijn notation'' of term graphs with labels \cite{Baren}.


\end{eRemark}


\section{Initial Algebra Semantics}\label{sec:ini}

In this section, we show that cyclic sharing terms form an initial
algebra and derive structural recursion and induction from it.

\subsection{Construction}
We use Fiore's approach to algebras for typed abstract syntax with 
binding \cite{NBE}  in 
the presheaf category $(\Set^{\FF\!\downarrow\! U})^U$ where $U$ is the set
of all types. 
Now, we take the set \TT of all shape trees for $U$, and 
the set \Nat of natural numbers for variables (i.e. pointers), 
instead of the category \FF of finite sets and all functions 
(used for renaming variables),
because we do not need renaming of pointers.

We define the discrete category $\FT$ by taking 
contexts $\Gamma = <\tau_1,\ooo,\tau_n>$ as objects
(which is equivalent to $\Nat\!\downarrow\!\TT$). 
We also regard \TT as a discrete category.
We consider algebras in \SetFTT.
Two preliminary definitions are required.
We define the presheaf $\PP \in \SetFT$ for pointers by
\begin{meqa}
\PP(<\tau_1,\ooo,\tau_n>)
  &= \set{ \ptr{p}{i} \| 1 \le i \le n,\; 
p \in \Pos(\tau_i)}.
\end{meqa}
For each $\tau \in \TT$,
we define the functor $\delta_\tau : \SetFT \rTo \SetFT$ 
for context extension by
$
\delta_\tau A = A(<\tau,->).
$

We define the \W{signature functor} $\Sig : \SetFTT \rTo \SetFTT$
\W{for cyclic sharing binary trees},
which takes $A\in \SetFTT$ and a type in \TT, and gives
a presheaf in \SetFT, as follows:

\begin{align*}
  (\Sig A)_{ \svoid} = 0 \qquad 
  (\Sig A)_{\sptr} = \PP \qquad
  (\Sig A)_{ \slf} = K_\ZZ \qquad 
  (\Sig A)_{  \sbin(\sig,\tau)} 
   = \delta_{ \sbin(\svoid,\svoid)} A_\sig \X \delta_{ \sbin(\sig,\svoid)} A_{\tau}
\end{align*}
\smallskip

\noindent
where $K_\ZZ$ is the constant functor to $\ZZ$, and $0$ is the empty
set functor.
A \W{\Sig-algebra} $A$ is a pair $(A,\alpha)$ consisting of 
a presheaf $A\in\SetFTT$ for a carrier and a natural transformation
$\alpha: \Sig A \to A$ for an algebra structure.
By definition of \Sig,
to give an algebra structure is to give the following 
morphisms of \SetFT:
\[
\fn{ptr}^A: \PP \to A_\sptr \qquad
  \lf^A: K_\ZZ \to A_{ \slf} \qquad
\bin^{\sig,\tau\, A}:  \delta_{\sbin(\svoid,\svoid)} A_\sig \X 
    \delta_{ \sbin(\sig,\svoid)} A_\tau \to 
 A_{ \sbin(\sig,\tau)}.
\]
A \W{homomorphism} \phi of \Sig-algebras 
from $(A,\alpha)$ to $(B,\beta)$
is a morphism $\phi:A \to B$
such that $\phi \o \alpha = \beta \o \Sig\phi$.

Let $T$ be the presheaf of all derivable cyclic sharing terms defined by
\[
T_\tau(\Gamma) = \set{t \| \Gamma \pr t :\tau}.
\]


\begin{theorem}\label{th:ini}
For the signature functor \Sig for cyclic sharing binary trees,
$T$ forms an initial \Sig-algebra.
\end{theorem}
\proof
Since $\delta_\tau$ preserves \omega-colimits, so does \Sig.
An initial \Sig-algebra is constructed by 
the colimit of the \omega-chain
$
0 \to \Sig 0 \to \Sig^2 0 \to  \ccc
$  \cite{Smyth-Plotkin}.
These construction steps correspond to
derivations of terms by typing rules, 
hence their union $T$ is the colimit.
The algebra structure $\fn{in} : \Sig T \to T$ 
of the initial algebra is obtained by one-step inference of the
typing rules, i.e.,
given by the following operations
\begin{meq}
\begin{array}[h]{rlrll}
\fn{ptr}^T(\Gamma): \PP(\Gamma) &\to T_\sptr(\Gamma) \qquad\quad&
  \lf^T(\Gamma): \ZZ &\to T_{ \slf}(\Gamma) \\
\ptr{p}{i} &\mapsto \ptr{p}{i}&
  k &\mapsto \lf( k)\\
\end{array}
\\
\bin^{T}(\Gamma):  T_\sig(\sbin(\svoid,\svoid), \Gamma) \X 
    T_\tau(\sbin(\sig,\svoid), \Gamma) \to 
 T_{ \sbin(\sig,\tau)}(\Gamma);\qquad \label{eq:ini-op}
s,\, t \mapsto \bin(s,t).  \nonumber
\end{meq}
\qed

The set $T_\tau(<>)$ is the set of all complete (i.e. no dangling pointers)
cyclic sharing trees of a shape $\tau$.

This development of an initial algebra characterisation
follows the line of \cite{FPT,NBE,typed-HOAS}.
Therefore, we can further 
develop a full theory of algebraic models of abstract syntax
for cyclic sharing structures
along this line.
It will provide
second-order typed abstract syntax with object/meta-level 
variables and substitutions via
a substitution monoidal structure 
and a free \Sig-monoid \cite{free,
Fiore2nd}
in \SetFTT
(by incorporating suitable arrows into \FT).
Object/meta-substitutions on cyclic sharing structures
will provide ways to construct
cyclic sharing structures from smaller structures in a sensible manner.
But this is not the main purpose of this paper.
Details will therefore be pursued elsewhere.

\subsection{Structural recursion principle}
An important benefit of initial algebra characterisation
is that the unique homomorphism from the initial to another algebra
is a mapping defined by structural recursion.

\begin{theorem}
The unique homomorphism \phi from the initial \Sig-algebra T to 
a \Sig-algebra $A$
is described as
\begin{meqa}
&\phi_\sptr(\Gamma)( \ptr{p}{i} ) = \fn{ptr}^A(\Gamma)(\ptr{p}{i})\\
&\phi_\slf(\Gamma)(\lf(k)) = \lf^A(\Gamma)(k)\\
&\phi_{\sbin(\sig,\tau)}(\Gamma)(\bin(s,t)) 
  = \bin^A(\Gamma) 
(\phi_\sig(\sbin(\svoid,\svoid), \Gamma)(s),\; 
 \phi_\tau(\sbin(\sig,\svoid), \Gamma)(t)).
\end{meqa}
\end{theorem}
\proof
Since the unique homomorphism
$\phi : T \rTo A$ is a morphism of \SetFTT.
\qed


\begin{eExample}\label{ex:height}
We give examples of functions on $T$
defined by structural recursion.
\begin{alignat*}{2}
\intertext{%
(i) The function \fn{leaves}
that collects all leaf values in a cyclic sharing tree
$t\in T_\tau(\Gamma)$:
}
&\fn{leaves} : T \rTo K_{\mathcal{P}(\ZZ)}\\
&\fn{leaves}_\sptr(\Gamma)( \ptr{p}{i} ) &=\;\;& \emptyset\\
&\fn{leaves}_\slf(\Gamma)(\lf(k)) &=\;\;& \set{k} \qquad\\
&\fn{leaves}_{\sbin(\sig,\tau)}(\Gamma)(\bin(s,t)) 
  &=\;\;& \fn{leaves}_\sig(\sbin(\svoid,\svoid), \Gamma )(s) 
    \union  
    \fn{leaves}_\tau(\sbin(\sig,\svoid), \Gamma )(t).\\
\intertext{%
This is because $\fn{leaves}$ is the unique homomorphism from $T$ to
a \Sig-algebra $K_{\mathcal{P}(\ZZ)}$ (the constant bifunctor to the
power set of integers \ZZ)
whose operations are given by
}
&\fn{ptr}^{K_{\mathcal{P}(\ZZ)}}(\Gamma)(\ptr{p}{i}) &=\;\;& \emptyset\\
&\lf^{K_{\mathcal{P}(\ZZ)}}(\Gamma)(k)  &=\;\;& \set{k}\\
&\bin^{K_{\mathcal{P}(\ZZ)}}(\Gamma)(x,y) &=\;\;& x \union y .\\
\intertext{%
(ii) The function \fn{height}
that computes the height of a cyclic sharing tree $t$:
}
&\fn{height} : T \rTo K_\ZZ\\
&\fn{height}_\sptr(\Gamma)( \ptr{p}{i} ) &=\;\;& 1\\
&\fn{height}_\slf(\Gamma)(\lf(k)) &=\;\;& 1 \qquad\\
&\fn{height}_{\sbin(\sig,\tau)}(\Gamma)(\bin(s,t)) 
  &=\;\;& \fn{max}(\fn{height}_\sig(\sbin(\svoid,\svoid), \Gamma)(s),
    \fn{height}_\tau(\sbin(\sig,\svoid), \Gamma )(t))+1\\
\intertext{%
where \fn{max} is the maximum function on \ZZ.
This is because \fn{height} is the unique homomorphism from $T$ to a \Sig-algebra
$K_\ZZ$
whose algebra structure is the obvious one.
Notice that the height is not so directly
defined in ordinary graph representations.
}
\intertext{%
(iii) The function \fn{skeleton}
that computes the shape of a given cyclic sharing tree $t$:
}
&\fn{skeleton} : T \rTo K_\TT\\
&\fn{skeleton}_\sptr(\Gamma)( \ptr{p}{i} ) &=\;\;& \sptr\\
&\fn{skeleton}_\slf(\Gamma)(\lf(k)) &=\;\;& \slf \qquad\\
&\fn{skeleton}_{\sbin(\sig,\tau)}(\Gamma)(\bin(s,t)) 
  &=\;\;& \sbin(\fn{skeleton}_\sig(\sbin(\svoid,\svoid), \Gamma)(s),\,\fn{skeleton}_\tau(\sbin(\sig,\svoid), \Gamma)(t)).
\end{alignat*}
\end{eExample}

From an algorithmic perspective,
the structural recursion principle provides
depth-first search traversal of a rooted graph.
Consequently, graph algorithms based on 
depth-first search are directly programmable
using this structural recursion.
On the author's home page \BR (\verb!http://www.cs.gunma-u.ac.jp/~hamana/!),
several other simple graph algorithms have been programmed using
structural recursion.

\subsection{Structural induction principle}
Another important benefit of initial algebra characterisation
is the tight connection to structural induction principle.
To derive it,
following \cite{Jacobs,JacobsBook},
we use the category \subSetFTT of predicates on \SetFTT defined by
\begin{enumerate}[$\bullet$]
\item objects: sub-presheaves $(P\incl U)$, i.e.,
inclusions between $P,U \in \SetFTT$,
\item arrows: $u:(Q\incl V) \to (P\incl U)$ 
are natural transformations $u:V\to U$ between underlying presheaves 
satisfying $a \in Q_\tau(\Gamma)$ implies $u(a) \in P_\tau(\Gamma)$
for all $\tau\in\TT, \Gamma\in\FT$.
\end{enumerate}\medskip

\noindent A sub-presheaf ($P\incl T$)
is seen as
a {predicate $P$ on cyclic sharing terms} $T$,
which is indexed by types and contexts.
So, we say
``$P_\tau^\Gamma(t)$ holds'' when $t\in P_\tau(\Gamma)$ for $t\in T_\tau(\Gamma)$.

We consider \Sig-algebras in \subSetFTT by 
``logical predicate'' lifting \cite{Jacobs}
of algebras in \SetFTT.
Why this is lifting is that now we consider
the functor $p : \subSetFTT \to \SetFTT$ sending
$(P \incl U)$ to the underlying presheaf $U$.
Then we lift the functor $\Sig$
to $\SigPred $ in a commuting diagram
\begin{diagram}[height=2em]
\subSetFTT & \rTo^\SigPred & \subSetFTT\\
\dTo^p     &               & \dTo_p\\      
\SetFTT    & \rTo_\Sig     & \SetFTT
\end{diagram}
by induction on the structure of \Sig:
\begin{meqa}
  &(\SigPred (P \incl U))_{\sptr} = (\PP \incl \PP)\\
  &(\SigPred (P \incl U))_{ \svoid} = (0\incl 0)\\
  &(\SigPred (P \incl U))_{ \slf} = (K_\ZZ \incl K_\ZZ)
  \\
  &(\SigPred (P \incl U))_{  \sbin(\sig,\tau)} 
   = \delta_{ \sbin(\svoid,\svoid)} (P\incl U)_\sig \X \delta_
  { \sbin(\sig,\svoid)} (P\incl U)_{\tau}
 \end{meqa}
where we also lift the context extension to
$\delta_\tau : \sub{\SetFT} \to \sub{\SetFT}$ defined by
$
\delta_\tau (A \incl B) = (A<\tau,-> \incl B<\tau,->).
$

A \SigPred-algebra structure
$\alpha:\SigPred (P\incl T) \to (P\incl T)$ can be
read as the induction steps in a proof by structural induction.
For example, the operation in \subSetFTT
\begin{meqa}
\bin^{\sig,\tau \, P}(\Gamma)&:  
(P_\sig(\sbin(\svoid,\svoid), \Gamma ) \incl T_\sig(\sbin(\svoid,\svoid),\Gamma))
\X 
(P_\tau(\sbin(\sig,\svoid), \Gamma) \incl T_\tau(\sbin(\sig,\svoid), \Gamma))
\\&
\to 
(P_{ \sbin(\sig,\tau)}(\Gamma) \incl T_{ \sbin(\sig,\tau)}(\Gamma))
\\
&s,t \mapsto \bin(s,t)
\end{meqa}
means that
``$\text{if }P_\sig^{\Gamma,\sbin(\svoid,\svoid)}(s) 
\;\&\; P_\tau^{\Gamma,\sbin(\sig,\svoid)}(t) \text{ holds,
 then }  P_{\sbin(\sig,\tau)}^\Gamma(\bin(s,t)) \text{ holds.}
$''

Jacobs showed that if
a fibration $\EE \to \BB$ satisfies several conditions
(having fibered (co)products, etc.),
then the logical predicate lifting from \BB to \EE
preserves initial algebras
(Prop. 9.2.7 in \cite{JacobsBook}).
The functor $p : \subSetFTT \to \SetFTT$ is actually such a fibration.
Consequently, because $T$ is an initial \Sig-algebra,
$(T \incl T$) is an initial \SigPred-algebra.
The unique homomorphism $\phi : (T \incl T) \rTo (P \incl T)$ means 
that $P$ holds for all cyclic sharing terms in $T$.
Hence

\begin{theorem}\label{th:str-ind}
Let $P$ be a predicate on $T$.
To prove that $P_\tau^\Gamma(t)$ holds for all $t \in T_\tau(\Gamma)$, it suffices to show
\begin{enumerate}[\em(i)]
\item $P_\sptr^\Gamma(\ptr{p}{i})$ holds for all $\ptr{p}{i}\in\PP(\Gamma)$,
\item $P_\slf^\Gamma(\lf(k))$ holds for all $k\in\ZZ$,
\item $\text{if }P_\sig^{\sbin(\svoid,\svoid), \Gamma}(s) 
\;\&\; P_\tau^{\sbin(\sig,\svoid), \Gamma}(t) \text{ holds,
 then }  P_{\sbin(\sig,\tau)}^\Gamma(\bin(s,t)) \text{ holds.}
$
\end{enumerate}
\end{theorem}

This structural induction principle is useful to prove 
properties of functions on cyclic sharing terms
defined by structural recursion.
As an example, we show the following simple property of the function
\fn{skeleton} defined in Example \ref{ex:height}.

\begin{proposition}
For all $t\in T_\tau(\Gamma)$, $\fn{skelton}_\tau(\Gamma)(t) = \tau$.
\end{proposition}  
\proof
By structural induction on $t$.
\begin{enumerate}[(i)]
\item Case $t =\ptr{p}{i}\in\PP(\Gamma)$.
By definition, $\fn{skelton}_\sptr(\Gamma)(\ptr{p}{i})=\sptr$.
\item Case $t=\lf(k)$.
By definition, $\fn{skelton}_\slf(\Gamma)(\lf(k))=\slf$.
\item Case $t=\bin(s_1,s_2)$. Then,
  \begin{meqa}
    \fn{skeleton}_{\sbin(\sig,\tau)}(\Gamma)(\bin(s_1,s_2)) 
   &=  \sbin(\fn{skeleton}_\sig(\sbin(\svoid,\svoid), \Gamma)(s_1),\fn{skeleton}_\tau(\sbin(\sig,\svoid), \Gamma)(s_2))\\
   &= \sbin(\sig,\tau) \qquad\text{by induction hypothesis.}
   \rlap{\hbox to94 pt{\hfill\qEd}}
  \end{meqa}
\end{enumerate}



\section{Inductive Types for Cyclic Sharing Structures}
\label{sec:impl}
In this section, we achieve
our goal [\ref{item:impl}] to give inductive types
for cyclic sharing structures.
We give implementations in two different systems.
We first use the functional language Haskell for an implementation
because
\begin{enumerate}[(i)]
\item we show that our characterisation of cyclic sharing is 
available in today's programming language technology, and
\item Haskell's type system is
powerful enough
to implement our initial algebra characterisation faithfully.
\end{enumerate}
Secondly, we give an implementation by dependent types
in the proof assistant Agda.

\subsection{A GADT definition in Haskell}\label{sec:has}

Because the set $T_\tau (\Gamma)$ of cyclic sharing terms
depends on a shape tree and context,
it should be implemented as a dependent type.
We have seen in the proof of Theorem \ref{th:ini} that
constructors of cyclic sharing terms are \TT and \FT-indexed
functions. Inductive types defined by indexed constructors have been known
as \W{inductive families} in dependent type theories
\cite{Dybjer94inductivefamilies}. 
Recently, the Glasgow Haskell Compiler (GHC) incorporates this feature
as \W{GADTs} (generalised algebraic data types) \cite{spj+:gadt}.
Using another feature called type classes, 
we can realise lightweight dependently-typed programming in 
Haskell \cite{fake}.

We will implement  $T_\tau (\Gamma)$ as
a GADT ``\code{T n t}'' that depends on a context \code{n} (for \Gamma)
and a shape tree \code{t} (for \tau). 
In Haskell, a type can only depend on types (not values).
For that reason, we firstly 
define type-level shape trees by using a type class.
{\small \begin{verbatim}
  data E
  data P
  data L     = StopLf
  data B a b = DnL a | DnR b | StopB 

  class Shape t
  instance Shape E
  instance Shape P 
  instance Shape L
  instance (Shape s, Shape t) => Shape (B s t)
\end{verbatim}}
\noindent
These define constructors 
of shape trees as types \code{E},\code{P},\code{L} and a type 
constructor \code{B}, then
group them by the 
type class \verb!Shape!.
Values of a shape tree type \tau
are defined by $\Pos(\tau)$, i.e. ``referable positions'' in \tau.
For example, consider a shape tree $\sbin(\sbin(\slf,\slf),\slf)$.
The position $1 \cdot 2$ in this shape tree
is coded as the well-typed term
\verb!DnL (DnR StopLF) :: B (B L L) L!
\noindent
where \code{StopLf} means stopping at a leaf.

Similarly, 
a typing context $<\tau_1,\ooo,\tau_n>$ is coded as a type-level sequence
\[
\code{TyCtx }\tau_1\; (\code{TyCtx }\tau_2 \ccc 
 (\code{TyCtx }\tau_n \code{ TyEmp}))
\]
and the type constructors
are grouped 
by the type class \verb!Ctx!.
Values of a context type 
are ``pointers'' (e.g. \verb!(Up UpStop)! meaning $\upptr{2}$). 
{\small \begin{verbatim}
  data TyEmp
  data TyCtx t n = Up n | UpStop | UpGD t

  class Ctx n
  instance Ctx TyEmp
  instance (Shape t, Ctx n) => Ctx (TyCtx t n)
\end{verbatim}}
\noindent
%
Finally, we define the set $T_\tau (\Gamma)$ as
a GADT ``\code{T}'' that takes a context and a shape tree as 
two arguments of the type constructor \code{T}.
{\small \begin{verbatim}
data T :: * -> * -> * where
  Ptr :: Ctx n => n -> T n P
  Lf  :: Ctx n => Int -> T n L
  Bin :: (Ctx n, Shape s, Shape t) =>
         T (TyCtx (B E E) n) s -> T (TyCtx (B s E) n) t -> T n (B s t)
\end{verbatim}}
\noindent
This defines three constructors of cyclic sharing terms faithfully.
Note that the part ``\verb!Ctx n =>!'',
called a context of a type class, is a
quantification meaning that ``for every type \verb!n! which is an instance of
the type class \verb!Ctx!''.

For example,
the term in Example \ref{ex:left} is certainly a well-typed term;
its type is inferred in the GHC (by invoking the command \verb!:t!
in the interpreter)
{\small \begin{verbatim}
  Bin (Bin (Lf 5) (Lf 6)) 
      (Bin (Ptr (Up (UpGD (DnL (DnL StopLf))))) (Lf 7))
  :: T TyEmp (B (B L L) (B P L))
\end{verbatim}}
\noindent The term \code{Up (UpGD (DnL (DnL StopLf)))} is the representation of
the pointer $\ptr{11}{2}$ in de Bruijn notation,
which is read from the top
as ``going up and up, then going down (\code{GD} is short for going down) to
the position $11$ and stopping at a leaf''.
The type inference and the type checker automatically
ensure well-formedness of cyclic sharing terms.

In Haskell, we can equally use
the GADT \code{T} as an ordinary algebraic datatype.
Therefore, we can define a function on it by structural recursion as 
described in Example \ref{ex:height} 
(even simpler; shape tree and context parameters
are unnecessary in defining functions because of Haskell's compilation 
method 
\cite{spj+:gadt}). 
The implementation and additional examples using the GADT \code{T} 
are available from the author's home page.

\subsection{A dependent type definition in Agda}

Secondly, we consider a definition in a proof 
assistant/dependently-typed programming
language Agda \cite{Agda}.
There are several ways for implementation.
One way is to use so-called universe construction \cite{powerPi}
by defining decoding functions
from type names to actual types to mimic the type class mechanism 
used in the previous subsection.
The resulting definition might resemble
the Haskell version.
Another way is more natural
to use the full power of dependent types
in Agda. 
In this subsection, we take this approach.
We implement the initial algebra $T$ of cyclic sharing tree structures
as a dependent type that depends on
two \W{values} (not types as in Haskell), a shape tree and a context.

We maximally use Agda's notational advantage, which allows
Unicode for mathematical symbols in a program.
In the following Agda code, we use mathematical symbols we have used 
in the paper to the greatest degree possible
(but it is 
certainly a real Agda code, not a pseudo-code).

We define shape trees as a usual inductive type, and contexts as
the type of sequences of shape trees (where $\send$ is the empty context and ``,'' 
is the separator):
\begin{haskell*}
\mydata Shape : Set \mywhere
{
   E &:\phantom{!}& Shape\\
   P &:& Shape\\
   L &:& Shape\\
   B &:& Shape \to Shape \to Shape
}
\end{haskell*}

\begin{haskell*}
\mydata Ctx : Set \mywhere
{
   \send &:\phantom{!}& Ctx\\
   \_ ,\_ &:& Shape \to Ctx \to Ctx\\
}
 \end{haskell*}
\noindent
The type \<Pos \tau\> for 
positions of a shape tree \tau is defined naturally as 
an Agda's inductive family, which consists
of indexed constructors.
The style of definition
is almost identical to that of GADTs in Haskell.

\newpage
\begin{haskell*}
\mydata Pos : Shape \to Set \mywhere
{
   \epsilon &:\phantom{!}& Pos L\\
   \epsilon' &:\phantom{!}& \forall \{\sig \tau\} \to Pos (B \sig \tau)\\
   DnL &:& \forall \{\sig \tau\} \to Pos \sig \to Pos (B \sig \tau)\\
   DnR &:& \forall \{\sig \tau\} \to Pos \tau \to Pos (B \sig \tau)\\
}
\end{haskell*}
\noindent
To define the dependent type $T$ for cyclic sharing trees,
the crucial ingredient is the implementation of the presheaf \PP for pointers,
recalling that it was defined by
\[
\PP(<\tau_1,\ooo,\tau_n>)
  = \set{ \ptr{p}{i} \| 1 \le i \le n,\; 
p \in \Pos(\tau_i)}.
\]
What we need is to implement a way to pick an index $i$ and 
$\tau_i$ from a typing context concisely.
The following type \<Index\> does this job.

\begin{haskell*}
\mydata Index : Shape \to Ctx \to Set \mywhere
{
  one  &:\phantom{!}& \forall \{\Gamma \tau\} \to Index \tau (\tau , \Gamma)\\
  s &:& \forall \{\Gamma \tau \sig\} \to Index \tau \Gamma  \to Index \tau (\sig ,\Gamma)
}
\end{haskell*}
\noindent
A well-typed term ``\<i : Index \tau \Gamma\>''   means
"$i$ is the index of a shape tree \tau in $\Gamma=\tau_1,\ooo,\tau,\ooo,\tau_n$",
e.g.,
\<s (s one) : Index \tau_3 (\tau_1 , \tau_2 , \tau_3 , \send)\>.
%
Then, the presheaf \PP is naturally implemented.
\begin{haskell*}
\mydata PO : Ctx \to Set \mywhere
{
   \ptr{\_}{\_} : \forall \{\Gamma \tau\} 
    \to Pos \tau \to Index \tau \Gamma \to PO(\Gamma)
}
\end{haskell*}
\noindent
Using these ingredients, the
implementation of the initial algebra $T$
for cyclic sharing trees is quite the same as
the mathematical definition we obtained in Theorem \ref{th:ini}.

\begin{haskell*}
\mydata T : Ctx \to Shape \to Set \mywhere
{
   ptr &:& \forall \{\Gamma\} \to PO(\Gamma) \to T \Gamma P\\
   lf &:\phantom{!}& \forall \{\Gamma\} \to Int \to T \Gamma  L\\
   bin &:& \forall \{\Gamma \sig \tau\} \to T (B E E , \Gamma) \sig
      \to T (B \sig E , \Gamma) \tau \to T \Gamma (B \sig \tau) \\
}
\end{haskell*}
\noindent
For example,
the term in Example \ref{ex:left} is a well-typed term 
also in Agda.
\begin{haskell*}
  bin (bin (lf 5) (lf 6)) 
      (bin (ptr (\ptr{\;DnL (DnL \epsilon)\;}{\;s one}) (lf 7)))
  : T \send (B (B L L) (B P L))
\end{haskell*}

\noindent Defining a function on the type $T$ by structural recursion
is directly possible
because of Agda's pattern matching mechanism on dependent types.
The functions in Example \ref{ex:height}
are defined directly.
In addition,
shape tree and context parameters can be (Agda's feature of)
implicit arguments. 
Therefore, we can use such functions concisely by omitting 
complex indices, as in the case of GADTs in Haskell.



\section{General Signature}\label{sec:gen}

We give construction of cyclic sharing structures for 
arbitrary signatures
as a natural generalisation of the binary tree case.

A \W{signature} \Sig for cyclic sharing structures
consists of a set \Sig of function symbols
having arities.
A function symbol of arity $n\in\Nat$ is
denoted by $f^{(n)}$.
Each function symbol $f$ has an associated \W{shape symbol} $\qu{f}$
(typically written in small caps such as \sbin).

\begin{eExample}
For the case of cyclic sharing binary trees,
the signature
\Sig consists of
$
\bin^{(2)}$ and $
{\lf}^{(0)}.
$
Corresponding shape symbols are defined by $\qu{\bin} = \sbin, \qu{\lf}=\slf$.
\end{eExample}

The set \TT of all shape trees is defined by
\[
\TT \ni \tau ::= \svoid \| \sptr \| \qu{f}(\tau_1,\ooo,\tau_n)
\quad \text{for each $f^{(n)}\in\Sig$}.
\]
The set of all contexts is 
\[
\FT= \set{ <\tau_1,\ooo,\tau_n> 
 \| n\in\Nat,\, i\in\set{1,\ooo,n},\, \tau_i \in\TT}.
\]
Positions are defined by 
\begin{meqa}
\Pos(\svoid) &= \Pos(\sptr) = \emptyset\\
\Pos(\sf(\tau_1,\ooo,\tau_n)) &= \set{\epsilon} \union
\set{1.p \| p \in \Pos(\tau_1)}\union \ooo \union
\set{n.p \| p \in \Pos(\tau_n)}.
\end{meqa}

\begin{framed}
\x{-1em}\textbf{Typing rules}
\begin{meq}
\ninfrule{Pointer}
{|\Gamma|=i-1 \infspc p \in \Pos(\sig)}
 {}
{\Gamma,\sig,\Gamma' \pr \ptr{p}{i} : \sptr}
\qquad\quad
\ninfrule{Function}
{\gamma_1, \Gamma  \pr t_1 : \tau_1 \infspc \ccc\infspc
 \gamma_n, \Gamma  \pr t_n : \tau_n \infspc f^{(n)}\in \Sig}
 {}
{\Gamma \pr f(t_1,\ooo,t_n) :   \sf(\tau_1,\ooo,\tau_n)}
\end{meq}
where $\gamma_1 = \sf(\svoid,\ooo,\svoid),\, 
\gamma_{i+1} = \sf(\tau_1,\ooo,\tau_i,\svoid,\ooo,\svoid)$
for each $1\le i \le n-1$. 
\end{framed}
\noindent
The shape trees $\gamma_i$'s are also used below.

This general case has the safety and uniqueness properties as well.

\begin{theorem}[Safety]\label{th:safe}
If a closed term $\pr t : \tau$ is derivable,
any pointer in $t$ points to a node in $t$.
\end{theorem}

\begin{theorem}[Uniqueness]
Given a rooted graph that is connected, directed and edge-ordered,
the term representation is unique.
\end{theorem}

Next we provide an initial algebra characterisation.
The base category is $\SetFTT$.
The presheaf $\PP$ of pointers is defined the same as in Sect. \ref{sec:ini}.
For a signature \Sig,
we associate a signature functor $\Sig : \SetFTT \to \SetFTT$ defined by
\[
  (\Sig A)_{ \svoid} = 0 \qquad 
  (\Sig A)_{\sptr} = \PP \qquad
  (\Sig A)_{  \sf(\tau_1,\ooo,\tau_n)} 
   = \prod_{1\le i \le n} \delta_{\gamma_i} A_{\tau_i}
\quad \text{for each $f^{(n)}\in\Sig$}.
\]

The following theorems are straightforward
generalisations of the corresponding theorems 
for the binary tree case;
hence proofs are straightforward.

\begin{theorem}[Initial algebra]
Let \Sig be a signature.
$T_\tau(\Gamma) = \set{t \| \Gamma \pr t :\tau}$ forms an initial
\Sig-algebra where operations are:
\[
\begin{array}[h]{lrlrl}
&\fn{ptr}^T(\Gamma): \PP(\Gamma) \;&\to T_\sptr(\Gamma)\qquad\qquad
&f^{T}(\Gamma):  \prod_{1\le i \le n} T_{\tau_i}(\gamma_i, \Gamma) \;&\to
 T_{ \sf(\tau_1,\ooo,\tau_n)}(\Gamma)\\
&\ptr{p}{i} \;&\mapsto \ptr{p}{i}
&t_1,\ooo, t_n \;&\mapsto f(t_1,\ooo,t_n). 
\end{array}
\]
\end{theorem}

\begin{theorem}[Structural recursion]
The unique homomorphism $\phi$ 
from the initial \Sig-algebra $T$ to a \Sig-algebra $A$ is described as
\begin{meqa}
&\phi_\sptr(\Gamma)( \ptr{p}{i} ) = \fn{ptr}^A(\Gamma)(\ptr{p}{i})\\
&\phi_{\sf(\tau_1,\ooo,\tau_n)}(\Gamma)(f(t_1,\ooo,t_n)) 
  = f^A(\Gamma) 
(\phi_{\tau_1}(\gamma_1, \Gamma)(t_1),\ooo,
\phi_{\tau_n}(\gamma_n, \Gamma)(t_n)).
\end{meqa}
\end{theorem}

\begin{theorem}[Structural induction]\label{th:str-ind}
To prove that $P_\tau^\Gamma(t)$ holds for all $t \in T_\tau(\Gamma)$, it suffices to show
\begin{enumerate}[\em(i)]
\item $P_\sptr^\Gamma(\ptr{p}{i})$ holds for all $\ptr{p}{i}\in\PP(\Gamma)$,
\item if $f^{(n)}\in\Sig$ and $P_{\tau_i}^{\gamma_i, \Gamma}(t_i)$
holds for all $i=1,\ooo,n$,
 then $P_{\sf(\tau_1,\ooo,\tau_n)}^\Gamma(f(t_1,\ooo,t_n))$ holds.
\end{enumerate}
\end{theorem}\medskip

\noindent Moreover, to give a GADT in Haskell and a dependent type in Agda
for cyclic sharing structures of a given signature
is straightforward, along the line of Sect. \ref{sec:impl} for the 
signature of binary cyclic sharing trees.



\section{Variations of the Form of Pointers}\label{sec:vari}

We have concentrated up to this point
on \W{unique representation} of
a given rooted graph.
This has been achieved by 
imposing the form of pointers in cyclic sharing trees
only \W{from right to left}.
In this section, we consider relaxation of this restriction as
a variation of the theme of the paper.

Actually, our algebraic framework, algebras in \SetFTT,
is not only for depth-first search trees.
Algebras in \SetFTT can model 
\W{trees with arbitrary pointers}, and
the form of pointers
can be controlled by types using shape trees. 
That is,
our framework has sufficient flexibility
to represent any form of pointers precisely.
In addition, from the application perspective,
other forms of pointers will be useful.
For example, one may need to invert pointers
in a cyclic sharing tree in some algorithms.
In such a case, one needs to use pointers \W{from left to right},
not only from right to left.

Syntactically, the form of pointers is determined by shape tree types of function symbols
and the definition of position function \Pos.
Consequently, relaxing the restrictions is an easy modifications
of the previous treatment.
Semantically, such variations of signature give
other algebras of functors in \SetFTT for 
trees with pointers of various forms.

\subsection{Left-to-right pointers}
\begin{figure}[t]
\begin{center}
\includegraphics[scale=.3]{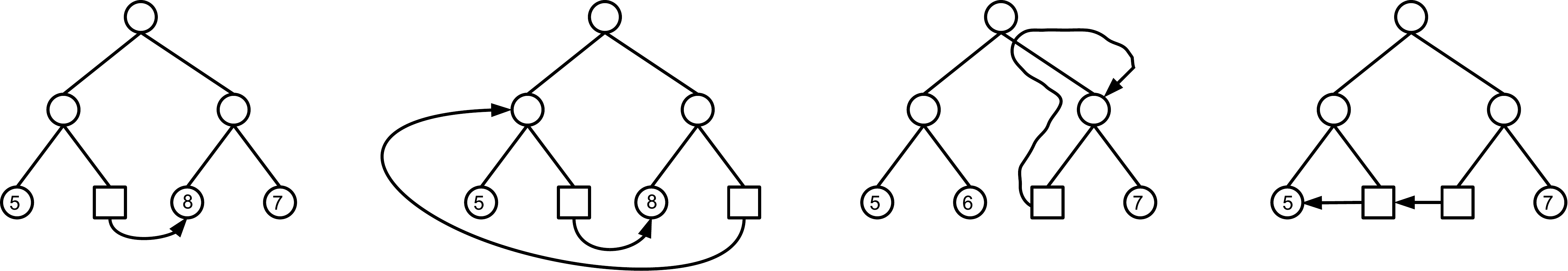}\\
{
\small
(i) Left-to-right 
\qquad
(ii) Both l-to-r \& r-to-l  \quad 
(iii) Allowing redundancy\quad
(vi) Indirect references
}
\caption{Trees involving various pointers}\label{fig:cs-vari}
\end{center}
\hrule height 0.5pt
\end{figure}

First, we consider cyclic sharing trees involving left-to-right pointers
and not involving right-to-left pointers.
An example is the tree (i) in Fig. \ref{fig:cs-vari},
as represented by
$\bin( \bin(\lf(5),\ptr{21}{2}),
        \bin(\lf(8), \lf(7)))$.
This case retains the uniqueness property of the representation for a given rooted graph.

A signature \Sig,
types \TT, contexts $\TT^*$ and positions \Pos are defined exactly the same
as they are in Sect. \ref{sec:gen}.

\begin{framed}
\x{-2em} \textbf{Typing rules}
\begin{meq}
\ninfrule{Pointer}
{|\Gamma|=i-1 \infspc p \in \Pos(\sig)}
 {}
{\Gamma,\sig,\Gamma' \pr \ptr{p}{i} : \sptr}
\qquad\quad
\ninfrule{Function}
{\gamma_1, \Gamma  \pr t_1 : \tau_1 \infspc \ccc\infspc
 \gamma_n, \Gamma  \pr t_n : \tau_n \infspc f^{(n)}\in \Sig}
 {}
{\Gamma \pr f(t_1,\ooo,t_n) :   \sf(\tau_1,\ooo,\tau_n)}
\end{meq}
where $\gamma_n = \sf(\svoid,\ooo,\svoid),\, 
\gamma_{i} = \sf(\svoid,\ooo,\svoid,\tau_{i+1},\ooo,\tau_n)$
for each $1\le i \le n-1$. 
\end{framed}

The category \SetFTT, a signature functor
$\Sig : \SetFTT \to \SetFTT$, and
the \Sig-initial algebra $T$ are also defined 
similarly to the definitions in Sect. \ref{sec:gen}.
Associated structural recursion and induction follow as well.

\begin{eExample}\label{ex:lr}
Consider the case of cyclic sharing binary trees involving
left-to-right pointers.
The difference from the original typing rules is the case of 
a binary node. Now, the above rule is instantiated as
\[
\ninfrule{Node}
{\sbin(\svoid,\tau), \Gamma  \pr s : \sig \infspc 
 \sbin(\svoid,\svoid), \Gamma  \pr t : \tau}
 {}
{\Gamma \pr \bin(s,t) :   \sbin(\sig,\tau)}
\]
This means that 
the shape tree type $\sbin(\svoid,\tau)$ at the left on the upper judgments
expresses that (a pointer in)
$s$ can point to a node in $t$,
whereas 
the shape tree type $\sbin(\svoid,\svoid)$ at the right
expresses
that (a pointer in)
$t$ cannot point to any node in $s$ by masking node information of $s$ by 
the void shape \svoid.
Actually, the general typing rule for left-to-right pointers
is obtained by generalising this observation.
\end{eExample}

\subsection{Symmetric form of pointers}
We can further allow
both right-to-left and left-to-right pointers.
An example is the tree (ii) in Fig. \ref{fig:cs-vari} represented by
$  \bin( \bin(\lf(5),\ptr{21}{2}),
        \bin(\lf(8), \ptr{1}{2})).
$
The only difference is the typing rules.

\begin{framed}
\x{-2em} \textbf{Typing rules}
\begin{meq}
\ninfrule{Pointer}
{|\Gamma|=i-1 \infspc p \in \Pos(\sig)}
 {}
{\Gamma,\sig,\Gamma' \pr \ptr{p}{i} : \sptr}
\qquad\quad
\ninfrule{Function}
{\gamma_1, \Gamma  \pr t_1 : \tau_1 \infspc \ccc\infspc
 \gamma_n, \Gamma  \pr t_n : \tau_n \infspc f^{(n)}\in \Sig}
 {}
{\Gamma \pr f(t_1,\ooo,t_n) :   \sf(\tau_1,\ooo,\tau_n)}
\end{meq}
where
$\gamma_{i} = \sf(\tau_1,\ooo,\tau_{i-1},\svoid,\tau_{i+1},\ooo,\tau_{n})$
for each $1\le i \le n$ (i.e. only $i$-th argument is set as \svoid).
\end{framed}

A shape tree
$\gamma_{i} = \sf(\tau_1,\ooo,\tau_{i-1},\svoid,\tau_{i+1},\ooo,\tau_{n})$
is used to prohibit only redundant reference 
(i.e. going up to an upper node and then going back down 
through the same path) by the void shape \svoid.

\begin{figure}[t]
\begin{center}
 \includegraphics[scale=.4]{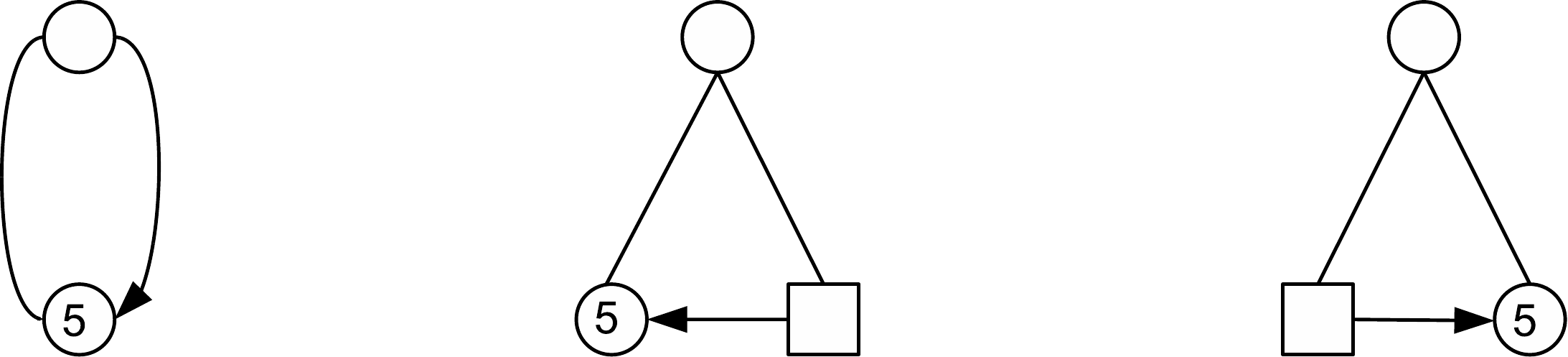}\\
\small
\qquad\qquad (i) A graph
\quad
(ii) Using right-to-left pointer\quad 
(iii) Using left-to-right pointer
\caption{Two representations of a graph}\label{fig:cs-xuniq}
\end{center}
\hrule height 0.5pt
\end{figure}

This case has no uniqueness property for a given graph,
because for example, a graph in Fig. \ref{fig:cs-xuniq} (i)
can be represented in two ways (ii) and (iii)
using cyclic sharing terms.

\subsection{No restriction of pointers}
In addition to the symmetric form of pointers, 
redundant references can be allowed.
An example is the tree (iii) in Fig. \ref{fig:cs-vari}
represented by $
  \bin( \bin(\lf(5),\lf(6)),
        \bin(\ptr{2}{2},\lf(7))).
$
Redundant reference means that 
the path obtained by $\ptr{p}{i}$ is not the shortest path to the
destination.
In the case of the tree (iii) in Fig. \ref{fig:cs-vari},
going up to the root and then going down 
through the same path to the right child.

\begin{framed}
\x{-2em} \textbf{Typing rules}
\begin{meq}
\ninfrule{Pointer}
{|\Gamma|=i-1 \infspc p \in \Pos(\sig)}
 {}
{\Gamma,\sig,\Gamma' \pr \ptr{p}{i} : \sptr}
\qquad\quad
\ninfrule{Function}
{\gamma, \Gamma  \pr t_1 : \tau_1 \infspc \ccc\infspc
 \gamma, \Gamma  \pr t_n : \tau_n \infspc f^{(n)}\in \Sig}
 {}
{\Gamma \pr f(t_1,\ooo,t_n) :   \sf(\tau_1,\ooo,\tau_n)}
\end{meq}
where $\gamma = \sf(\tau_1,\ooo,\tau_n)$.
\end{framed}

\subsection{Allowing indirect references}
Up to this point in the discussion,
we have assumed that a pointer cannot point to 
another pointer node.
Like the tree (vi) in Fig. \ref{fig:cs-vari} has been prohibited
because we aimed to obtain a unique
representation for a graph.
However, that assumption can also be relaxed.
This is achieved by merely modifying the definition of \Pos as
\[
\Pos(\sptr) = \set{\epsilon}.
\]
The tree (vi) in Fig. \ref{fig:cs-vari} is represented by
$
  \bin( \bin(\lf(5),\ptr{1}{1}),
        \bin(\ptr{12}{2},\lf(7))).
$

\subsection{Pointers from inner nodes}
We can allow pointers from inner nodes, i.e., not only
from leaves as we have considered.
This is by introducing a new term construct
\[
f(\ptr{p}{i} ; t_1,\ooo,t_n) 
\]
which expresses that an inner node $f$ having $n$-children also
has a pointer slot.
Typing rule is the combination of the previous term formations for
the pointer and function term.

\begin{framed}
\x{-2em} \textbf{Typing rule}
\begin{meq}
\ninfrule{Function}
{|\Gamma|=i-1 \infspc p \in \Pos(\sig) \infspc
\gamma, \Gamma  \pr t_1 : \tau_1 \infspc \ccc\infspc
 \gamma, \Gamma  \pr t_n : \tau_n \infspc f^{(n)}\in \Sig}
 {}
{\Gamma,\sig,\Gamma' \pr f(\ptr{p}{i} ; t_1,\ooo,t_n) :   \sf(\tau_1,\ooo,\tau_n)}
\end{meq}
where
$\gamma_{i} = \sf(\tau_1,\ooo,\tau_{i-1},\svoid,\tau_{i+1},\ooo,\tau_{n})$
for each $1\le i \le n$.
\end{framed}

This form of terms is used as a data model of XML called
\W{trees with pointers} \cite{contextlogic} by
Calcagno, Gardner and Zarfaty.

\subsection{Mixing variations}
The form of pointers need not be uniform 
(i.e. all pointers must be the same form)
as described above.
For example, in a single tree,
it is possible that some function symbols allow
only right-to-left pointers, some others allow only left-to-right, and
some others allow both, etc.
This possibility is realised merely by assigning an appropriate type to 
each function symbol, which
shows that our framework has 
expressive power to control the form of pointers.

It is important to note that
in any variation of typing rules,
the safety property of pointers stated in Theorem \ref{th:safe}
still holds.
Truly dangling
pointers cannot happen in this framework.


\section{Connection to Equational Term Graphs in the Initial Algebra Framework}
\label{sec:etg}

We have investigated a term syntax for cyclic sharing structures,
which gives a representation of a graph.
In this section, we give the converse, i.e., an explicit way to calculate
the graph for a given cyclic sharing term.
This means to give {semantics} of a cyclic sharing term by a
finite graph.
We give it using
Ariola and Klop's equational term graphs in the initial algebra framework.
This semantics also clarifies connections to existing works 
that have explored the semantics
of cyclic sharing structures. 

\begin{wrapfigure}[10]{l}{10em}
  \centering
\includegraphics[scale=.3]{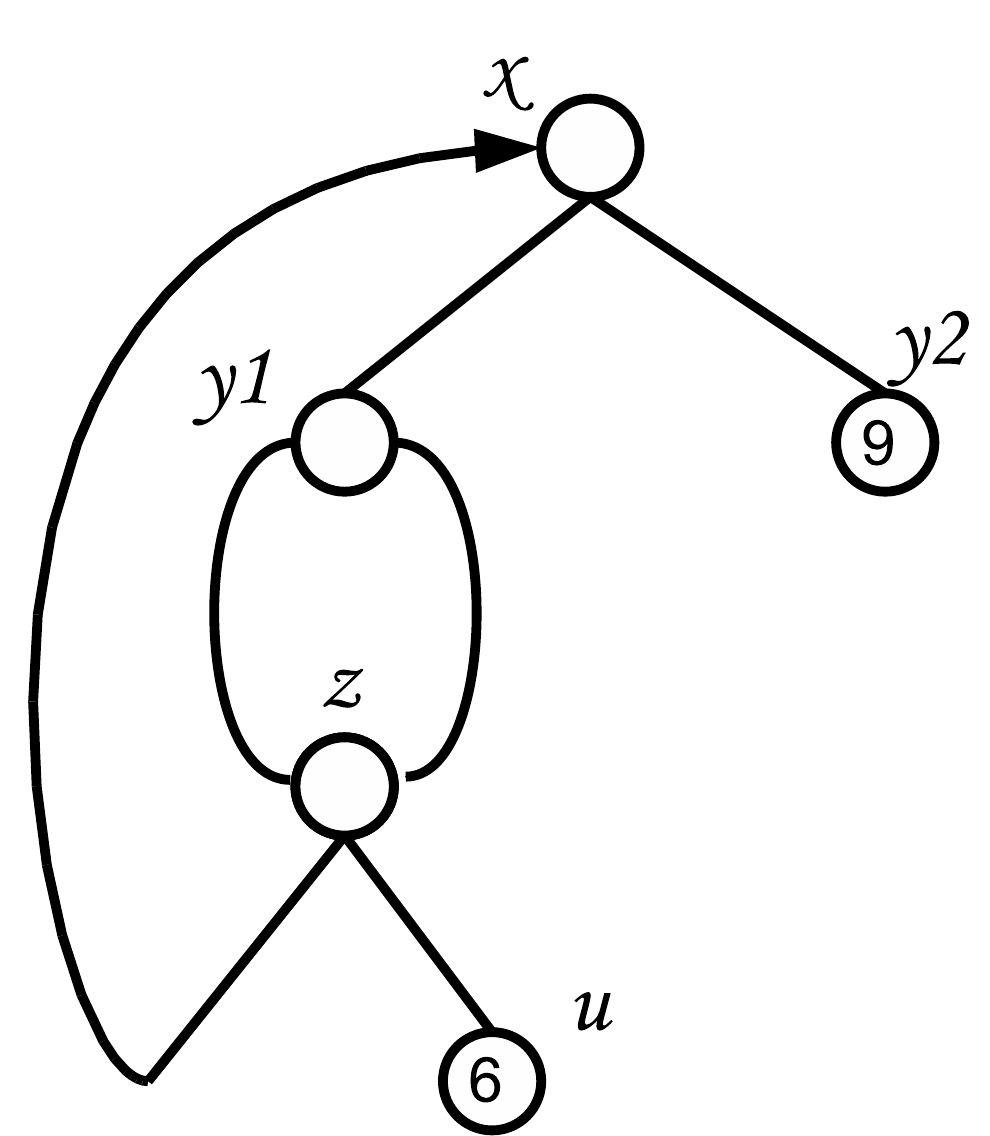}
\caption{rooted graph}\label{fig:clis4}
\end{wrapfigure}

Equational term graphs \cite{ETGRS} are another representation of 
cyclic sharing trees, which have been used in a formulation
of term graph rewriting.
This is a representation of a rooted
graph\footnote{%
For the case of an unrooted graph, 
it can be rooted by choosing an arbitrary
starting node. 
It may also have several other distinct connected components, which
might be represented by a set of equational term graphs.
}
by associating a unique name to each
node and
by writing down the interconnections through a set of recursive equations.
For example, the graph portrayed in Figure \ref{fig:clis4}
is represented as an equational term graph 
\[
\begin{array}[h]{lllllll}
\set{x \| &x &=\bin(y_1,y_2), \quad 
  &y_2 &= \lf(9),\\
  &y_1 &= \bin(z,z),
  &z &= \bin(x,u),\\
  &u &= \lf(6)
}.
\end{array}
\]
%
%

We use this form of equational term graphs, 
which is called flattened form
in \cite{ETGRS}, and which is formally defined as follows 
(NB. it differs slightly 
from the original syntax to make explicit the connection to cyclic sharing terms).

Suppose a signature \Sig and
a set $X=\set{x,x_1,\ooo}$ of variables.
\W{An equational term graph} is of the form
\[
\set{x \| x_1=t_1,\, x_2=t_2,\ooo}
\]
where each $t_i$ follows the syntax
\[
t :: = x \| \ptr{p}{i} \| f(x_1,\ooo,x_n).
\]
A variable is called \W{bound}
if it appears in the left-hand side of an equation; 
it is called \W{free} otherwise.
We also call $\ptr{p}{i}\;$ a
\W{free variable} (and regard it as a free variable).
We assume that any useless equation $y=t$, where $y$ cannot be reachable from
the root, is automatically removed in the presentation of equational term 
graphs \cite{ETGRS} (hence, equational term graphs are always connected and single-rooted).

%
We define a 
translation from a cyclic sharing term to an equational term graph
by the unique homomorphism from the initial algebra to 
an algebra consisting of equational term graphs.
The idea is to use positions as unique variables in an equational term graph.
We define
$\ETG_\tau(\Gamma)$ by the set of all equational term graphs 
having free variables taken from $\PP(\Gamma)$
(where a shape index \tau is meaningless for equational term graphs, 
but we just put this index to form a presheaf).
This \ETG forms a presheaf in \SetFTT.
Any equational term graph can be drawn as a tree-like graph
as Fig. \ref{fig:clis4} by traversing each node
in a depth-first search manner from the root.
Therefore, we can assign each node to its position in the whole equational term graph.
Consequently, an equational term graph
\[\set{x_1 \| x_1=t_1,\, x_2=t_2,\ooo}\]
can be normalised to an ``\alpha-normal form'' in which 
for each $x = t$,
the bound
variable $x$ is renamed to the position of $t$ in the whole term as
\[
\set{\epsilon \| \epsilon = t_1',\, 1=t_2',\ooo}  
\]
(see Example \ref{ex:transEgraph}).
We identify an equation term graph with its \alpha-normal form.

\begin{proposition}
\ETG forms a \Sig-algebra. 
The unique homomorphism
$
\den{-} : T \rTo \ETG
$
is monomorphic,
giving
an interpretation of a cyclic sharing term as
a graph represented by an equational term graph.
\end{proposition}
\proof
We define an algebra structure on \ETG of \alpha-normal forms as 
follows.
\begin{meqa}
& f^\ETG_\tau(\Gamma)(\set{\epsilon \| G_1},\ooo, \set{\epsilon \| G_n})
= \set{\epsilon \| \epsilon = f(1,\ooo,n),\,
 G'_1,\ooo, G'_n } \\
&\qquad \WHERE \set{1 \| G'_1}  =  \fn{shift}_1(\set{\epsilon \| G_1}) \;\ccc\;
\set{n \| G'_n}  =  \fn{shift}_n(\set{\epsilon \| G_n})
\nonumber\\
& \fn{ptr}^\ETG_\tau(\Gamma)(\ptr{p}{i}) = \set{\epsilon \| \epsilon=\ptr{p}{i}}
\\[.5em]
&\fn{shift}_i \set{\epsilon \| \epsilon = t_1,\; 1 = t_2, \ooo}
=  \set{i \| i = \fn{shift}_i(t_1),\; i.1= \fn{shift}_{i}(t_2), \ooo}
\\
&\fn{shift}_i(p) = i.p \quad \text{for a position } p
\\
&\fn{shift}_i(f(x_1,\ooo,x_n)) 
 = f(i . x_1,\ooo, i . x_n)\\
&\fn{shift}_i(\ptr{p}{x}) = \bigcur{
\ptr{p}{x-1} & \IF x\gt 1 \\
p            & \IF x = 1 
}
\end{meqa}
The function $\fn{shift}_i$ shifts every bound variable by a position $i\in\Nat$
(i.e. appending $i$ as prefix)
in a term to form an equational term graph suitably. 
Then, it is obvious that $\den{-}$ is monomorphic and
that it gives a translation from cyclic sharing terms to equational term graphs.
\qed

Notice that $\den{-} : T \rTo \ETG$ is not an isomorphism. 
Equational term graphs have much more freedom to express graphs
than cyclic sharing terms. For example, although
$\set{x \| x = x}$
is a valid equational term graph (the ``black hole''),
no corresponding cyclic sharing term exists.

\begin{eExample}\label{ex:transEgraph}
Consider the term
$
\mu x.\bin(\mu y_1.\bin(\mu z.\bin(\upptr x,\lf(6)),\ptr{1}{y_1}),\lf(9))$
of Fig. \ref{fig:cstree4}. This is represented as the following
term in de Bruijn
and is interpreted as an equational term graph:
\[
\begin{array}[h]{clll}
& 
  \bin(\bin(\bin(\upptr 3,\lf(6)),\ptr{1}{1}),\;\lf(9))\\[.5em]

\stackrel{\den{-}}{\mapsto} &
  \begin{array}[h]{llllllll}
  \set{\epsilon \| &\epsilon &= \bin(1,2),\qquad
                    & 12      &= \cross{11} ,\qquad
                     & 112     &= \lf(6),\\
                   &       1 &= \bin(11,12),\qquad
                    & 111     &= \cross{\epsilon}  ,\qquad
                     & 2       &= \lf(9),\\
                   & 11      &= \bin(111,112)
  }.
  \end{array}
\end{array}
\]
\end{eExample}




\section{Further Connections to Other Works}\label{sec:othersem}

The semantics of cyclic sharing terms by equational term graphs opens 
connections to other semantics
as $T \rTo \ETG \rTo \SS$, where \SS is any of the 
following semantics of
equational term graphs.

\begin{enumerate}[(i)]
\item \sLETREC-expressions: an equational term graph is
obviously seen as a \sLETREC-expression%
\footnote{%
\sLETREC-expressions are more expressive than equational term graphs
because they can express multiple roots by putting a tuple $(x_1,\ooo,x_n)$
of roots of distinct connected components 
in the body of a \sLETREC-expression \cite{Hassei}. 
}.\label{item:letrec}
\item Domain-theoretic semantics: mentioned below. \label{item:domain}
\item Categorical semantics in terms of traced symmetric monoidal
categories \cite{Hassei}. \label{item:traced}
\item Coalgebraic semantics: a graph is regarded as a coalgebraic structure
that produces every node information along its edges, 
e.g. \cite{AczelAdamek}. \label{item:coalg}
\end{enumerate}\medskip

\noindent The domain-theoretic semantics of \sLETREC-expressions
or systems of recursive equations (e.g. \cite{Courcelle}), is now
standard; it gives infinite expansion of cyclic sharing structures.
Via equational term graphs, we can interpret our cyclic sharing
terms in each of these semantics.
Each semantics has its own advantage and principles
related to some aspects of cyclic sharing structures.
However, \textit{none of these has focused on
our goals}, which are the following.
[\ref{item:ind}] A simple term syntax that admits 
structural induction.
[\ref{item:impl}] Direct usability in functional programming,
as described in the Introduction. Therefore, we have chosen the
initial algebra approach to cyclic sharing structures.

Although insufficient,
the above semantics (\ref{item:letrec}) and (\ref{item:domain}) are close
to our goals in the following way.
Consider the cyclic sharing term
$
\mu x.\bin(\mu y_1.\bin(\mu z.\bin(\upptr x,\lf(6)),\ptr{1}{y_1}),\lf(9))$
of Fig. \ref{fig:cstree4}.
As considered in Example \ref{ex:transEgraph},
this is interpreted as an equational term graph:
\[
\begin{array}[h]{clll}
  \begin{array}[h]{llllllll}
  \set{\epsilon \| &\epsilon &= \bin(1,2),\qquad
                    & 12      &= \cross{11} ,\qquad
                     & 112     &= \lf(6)\\
                   &       1 &= \bin(11,12),\qquad
                    & 111     &= \cross{\epsilon}  ,\qquad
                     & 2       &= \lf(9)\\
                   & 11      &= \bin(111,112)
  }.
  \end{array}
\end{array}
\]
Using domain-theoretic semantics, we can obtain its expansion
as an infinite term
\begin{equation}
\begin{array}[h]{llll}
\bin( \bin( \termlet ,\;\;
            \termlet ),\; \lf(9))
\end{array}
\label{eq:expansion}
\end{equation}
where each ``$\ccc$'' is actually an infinitary long that repeats
the whole term.
This is regarded as an expansion of the structure in which
each pointer node ``$\ptr{p}{i}$'' is connected directly to
the referred node.

Defining this idea in a lazy functional language
based on domain-theoretic semantics
such as Haskell
yields another interesting representation 
related to the use of internal pointer
structures.
Let's consider this in Haskell.
Let the type $\CST$ be a lazy datatype of trees defined by
\[
t ::= \lf(k) \| \bin(t_1,t_2)
\]
(but here, for simplicity,
we retain mathematical notation rather than Haskell).
Consequently, we define
the translation function
$\trc : \ETG \rTo \CST$
from equational term graphs to \CST by
\begin{equation}
\trc( \set{y_1 \| y_1 = r_1,\ooo, y_n = r_n})
\;=\;\; \LET (\vec x) = (\vec r )[\vec y \mapsto \vec x] \IN x_1
\label{eq:trc}
\end{equation}
where vectors denote sequences, 
and $[- \mapsto -]$ a substitution function of variables 
(written in Haskell).
At the level of Haskell,
this gives a translation into
{internal pointer structures} in the heap memory of an implementation,
because a \sLET-expression (which is theoretically \sLETREC)
generates a pointer structure
as presented in Fig. \ref{fig:clis4} 
because of the graph reduction mechanism of Haskell.
Printing it will generate an infinite term as Eq.
(\ref{eq:expansion}).
In this way, starting from $T$ via equational term graphs,
our cyclic sharing terms
can be used as ``blueprints'' of
pointer structures in the memory.

A problem in the pointer structures is
lack of structural induction.
Exactly how it is possible to compose and decompose
the pointer structures cleanly at the level of Haskell programming remains unclear.
Therefore, this approach was thought to be somewhat
insufficient for our goals, but 
this approach is nevertheless efficient and interesting.


\section{Conclusion}
We have given an initial algebra characterisation of cyclic sharing
structures and derived inductive datatypes, structural recursion, and 
structural induction on them. 
We have also associated them with
equational term graphs in the initial algebra framework. 
Hence
we have shown that various ordinary semantics of cyclic sharing structures
are applied equally to them.

From a programming perspective, practicality of our datatype
of cyclic sharing structures
must still be investigated.
A possible direction of future work is to seriously use
a dependently-typed programming language such as Coq and Agda
for programming with cyclic sharing structures
as an extension of this work.


\section*{Acknowledgement}
The basis for this work, which was motivated by a question
of Zhenjiang Hu, was done while the author visited IPL, University of Tokyo
during April 2007 - March 2008. I express my sincere gratitude to
Masato Takeichi and members of IPL for the opportunity to stay
in a stimulating and pleasant research environment.
I am also grateful to Varmo Vene and Tarmo Uustalu for 
discussions about datatypes in Haskell and reading early draft, and
Shin-Cheng Mu for a discussion about dependent types in Agda.
This work is supported by
the JSPS Grant-in-Aid for Scientific Research (19700006) and
NII collaboration research grant.



\bibliographystyle{alpha}
\bibliography{bib}

\end{document}